\documentclass[prd,superscriptaddress,twocolumn,preprintnumbers,showpacs,nofootinbib]{revtex4-1} 
\pdfoutput=1
\usepackage{amssymb,amsmath,graphicx,epsfig}
\usepackage{hyperref,tabu,color,array}
\usepackage{float,url}
\usepackage{hyperref}
\usepackage{slashed}
\usepackage{multirow,bm}
\usepackage[titletoc,title]{appendix} 
\usepackage[export]{adjustbox}
\usepackage[utf8]{inputenc}
\usepackage{xcolor}
\usepackage[normalem]{ulem}
\hypersetup{linktocpage=true}
\hypersetup{
colorlinks=true,   % false:	 links; true:colored links
linkcolor=blue,    % color of internal links
citecolor=blue,    % color of links to bibliography
filecolor=blue,     % color of file links
urlcolor=blue      % color of external links
}
\def\beq{\begin{equation}} 
\def\eeq{\end{equation}} 
\def\bea{\begin{eqnarray}} 
\def\eea{\end{eqnarray}} 
\def\nn{\nonumber}

\def\cvll{\rm C_V^{LL}}
\def\cvrl{\rm C_V^{RL}}
\def\csll{\rm C_S^{LL}}
\def\csrl{\rm C_S^{RL}}
\def\ctll{\rm C_T^{LL}}
\def\cvrr{\rm C_V^{RR}}
\def\Jps{{J/\psi}}
\def\bal#1\eal{\begin{align}#1\end{align}}
\begin{document} 
\title{B-meson charged current anomalies: the post-Moriond status}
%%%%%%%%%%%%%%%%%%%%%%%%%%%%%%%%%%%%%%%%%%%%%%%%%%%%%%%%%%%%%%%%%%%%%%%%%%%%%%% 
\author{Debjyoti Bardhan} 
\email{bardhan@post.bgu.ac.il} 
\affiliation{\normalfont{Department of Physics, Ben-Gurion University, Beer-Sheva 8410501, Israel}}
\author{Diptimoy Ghosh}
\email{diptimoy.ghosh@iiserpune.ac.in} 
\affiliation{\normalfont{Department of Physics, Indian Institute of Science Education and Research, Pune 411008, India}}
%%%%%%%%%%%%%%%%%%%%%%%%%%%%%%%%%%%%%%%%%%%%%%%%%%%%%%%%%%%%%%%%%%%%%%%%%%%%%%% 
\begin{abstract} 
In this note, we discuss the impact of the recent Belle result on the various theoretical explanations of the $R_D$ and $R_{D^*}$ anomalies.
The pure tensor explanation, which was strongly disfavoured by the measurements of $F_L^{D^*}$ and high-$p_T$ $p \, p \to \tau \, \nu$ searches
before Moriond, is now completely allowed because of reduction of the experimental world-average. Moreover, the pure right-chiral vector solution
(involving right-chiral neutrinos) has now moved  into the $2\sigma$ allowed range of the LHC $p \, p \to \tau \, \nu$ searches.
We also critically re-examine the bound on $\mathcal{B}(B_c^- \to \tau^- \bar{\nu}_\tau)$ from LEP data and show that the bound is considerably
weaker than the number $10\%$ often used in the recent literature. 
\end{abstract}
%%%%%%%%%%%%%%%%%%%%%%%%%%%%%%%%%%%%%%%%%%%%%%%%%%%%%%%%%%%%%%%%%%%%%%%%%%%%%%%
%\keywords{}
%\pacs{14.40.Nd, 13.20.He, 12.60.Cn}
%\preprint{}
\maketitle
%14.40.Nd   Bottom mesons (|B| > 0) 
%13.20.He   Decays of bottom mesons 
%12.60.Cn   Extensions of electroweak gauge sector
%%%%%%%%%%%%%%%%%%%%%%%%%%%%%%%%%%%%%%%%%%%%%%%%%%%%%%%%%%%%%%%%%%%%%%%%%%%%%%%
%\subsection{\rm \bf Introduction}
%\vspace*{-3mm}
%%%%%%%%%%%%%%%%%%%%%%%%%%%%%%%%%%%%%%%%%%%%%%%%%%%%%%%%%%%%%%%%%%%%%%%%%%%%%%%
%
%
The Belle collaboration has recently published results for $R_D$ and $R_{D^*}$ with a semileptonic tag \cite{Belle2019, Abdesselam:2019dgh}, and their result is consistent
with the Standard Model (SM) expectation within $1.2\sigma$.  Consequently, the experimental world average has moved towards the SM.
However, the tension between the experimental world average and the SM expectation is still more than $3\sigma$, and thus, it is interesting
to re-examine the status of the various New Physics (NP) explanations in view of the new world-average.  In Table.~\ref{tab-exp} below, we
collect all the experimental results related to this anomaly. 
%
%%%%%%%%%%%%%%%%%%%%%%%%%%%
\begin{table}[h!]
\begin{center}
\begin{tabular}{|p{0.7cm}|p{1.75cm}c|p{3.85cm}c|}
\hline 
 & \multicolumn{2}{c|}{SM prediction}  & \multicolumn{2}{c|}{Measurement}  \\ 
\hline
\multirow{ 2}{*}{$R_D$}    &  $0.300 \pm 0.008$ &\cite{Aoki:2016frl}      & $0.407 \pm 0.046$ (pre-Moriond) & \cite{Amhis:2016xyh} \\
                 &  $0.299 \pm 0.011$ & \cite{Na:2015kha}     & $0.334 \pm 0.031$                      & \cite{Amhis:2016xyh,Belle2019,Abdesselam:2019dgh} \\
\hline
\multirow{ 2}{*}{$R_{D^*}$}    &  \multirow{ 2}{*}{$0.258 \pm 0.005$} & \multirow{ 2}{*}{\cite{Bigi:2017jbd,Jaiswal:2017rve,Bernlochner:2017jka,Amhis:2016xyh}} & $0.306 \pm 0.015$ (pre-Moriond) & \cite{Amhis:2016xyh} \\
                         &                                 &                                & $0.297 \pm 0.015$                      &  \cite{Amhis:2016xyh,Belle2019}    \\
\hline
$P_\tau^{D^*}$ & $-0.47 \pm 0.04$     &\cite{Bigi:2017jbd}   &    $-0.38^{+0.55}_{-0.53}$   
&  \cite{Hirose:2016wfn,Hirose:2017dxl}                 \\
\hline
$F_L^{D^*}$ &  $0.46 \pm 0.04$           &       & $0.60 \pm 0.087$ & \cite{Abdesselam:2019wbt} \\
\hline
$R_{\Jps}$ &  $  0 .290 $  &     & $0.71 \pm 0.25$ & \cite{Aaij:2017tyk} \\
\hline
\end{tabular}
\caption{\sf Observables, their SM predictions and the experimentally measured values. The pre-Moriond experimental averages for $R_D$ and 
$R_{D^*}$ are based on \cite{Lees:2012xj, Lees:2013uzd, Huschle:2015rga, Aaij:2015yra, Sato:2016svk, Hirose:2016wfn, Hirose:2017dxl, Aaij:2017deq, Aaij:2017uff}.
%The SM prediction of $R_{\Jps}$ is based on the form-factors given in \cite{Wen-Fei:2013uea}. As the $B_c \to J/\psi$ form-factors  are not very
%reliably known, we do not show any uncertainty for $R_{\Jps}$. However, it is expected to be similar to that of $R_{D^\ast}$.
\label{tab-exp}}
\end{center} 
\end{table}
%%%%%%%%%%%%%%%%%%%%%%%%%%%%%%%%%%%%%%%%%%%%%%%%%%%%%%%%%%%%%%%%%%%% 
%

The most general effective Lagrangian for the decay $b \to c \, \tau^- \, \bar{\nu}_\tau$ involving mass dimension-6 operators and
only left-chiral neutrinos can be written as
\begin{align}
& {\cal L}^{b \to c \, \tau \, \nu}_{\rm eff} =  
-\frac{4 G_F V_{cb}}{\sqrt{2}} \left(  \cvll  \,    [\bar{c} \, \gamma^\mu P_L \, b] [\bar \tau \, \gamma_\mu \, P_L \, \nu]  \right.  \nn \\
                                         + &  \left.  \cvrl \,  [\bar{c} \, \gamma^\mu P_R \, b] [\bar \tau \, \gamma_\mu \, P_L \, \nu]  
                                                      + \csll [\bar{c} \, P_L \, b] [\bar \tau \, P_L \, \nu]  \right. \label{eff-lag} \\ 
                                         + &  \left. \csrl [\bar{c} \, P_R \, b] [\bar \tau \, P_L \, \nu] 
                                                      + \ctll [\bar{c} \, \sigma_{\mu \nu} P_L \, b] [\bar \tau \, \sigma^{\mu \nu} \, P_L \, \nu]  + \rm h.c. \right) \nn
\end{align} 
If one uses power-counting rules arising from linearly-realised $\rm SU(2) \times U(1)$ gauge invariance, it turns out that
the Wilson Coefficient (WC) $\cvrl$, with the possibility of lepton non-universality, is only generated at the mass dimension-8 level \cite{Azatov:2018knx}.
Thus, it is expected to be suppressed compared to the other WCs as long as the scale of NP is not too close to the Higgs vacuum expectation value,
thus we will ignore it in this analysis.

If one also assumes the existence of light right-chiral neutrino(s), as was first done in \cite{Becirevic:2016yqi} to solve the $R_D$ anomaly, five additional operators can be constructed by the replacement
$P_L \to P_R$ in the leptonic currents of Eq.~\ref{eff-lag}. In particular, a pure-right chiral vector current namely, 
\bal
 \hspace*{-2.5mm}{\cal L}^{b \to c \, \tau \, \nu}_{\rm eff} \supset  -\frac{4 G_F V_{cb}}{\sqrt{2}}
 \cvrr  \,    [\bar{c} \, \gamma^\mu P_R \, b] [\bar \tau \, \gamma_\mu \, P_R \, \nu] + \rm h.c. \, 
\eal
 was considered by several authors \cite{Asadi:2018wea, Greljo:2018ogz, Azatov:2018kzb} , and we will include it in our analysis.

As the experimental situation for $R_D$ and $R_{D^*}$ is far from clear, we do not try to perform a fit to the WCs; for an early global fit,
see \cite{Freytsis:2015qca}. Instead, we show how
$R_D$ and $R_{D^*}$ vary with respect to the WCs, and overlay the current $1\sigma$ experimental world-average and the corresponding
currently allowed values of the WCs.

\begin{figure}[!h!]
\centering
\begin{tabular}{cc}
\hspace*{-5mm}\includegraphics[scale=0.32]{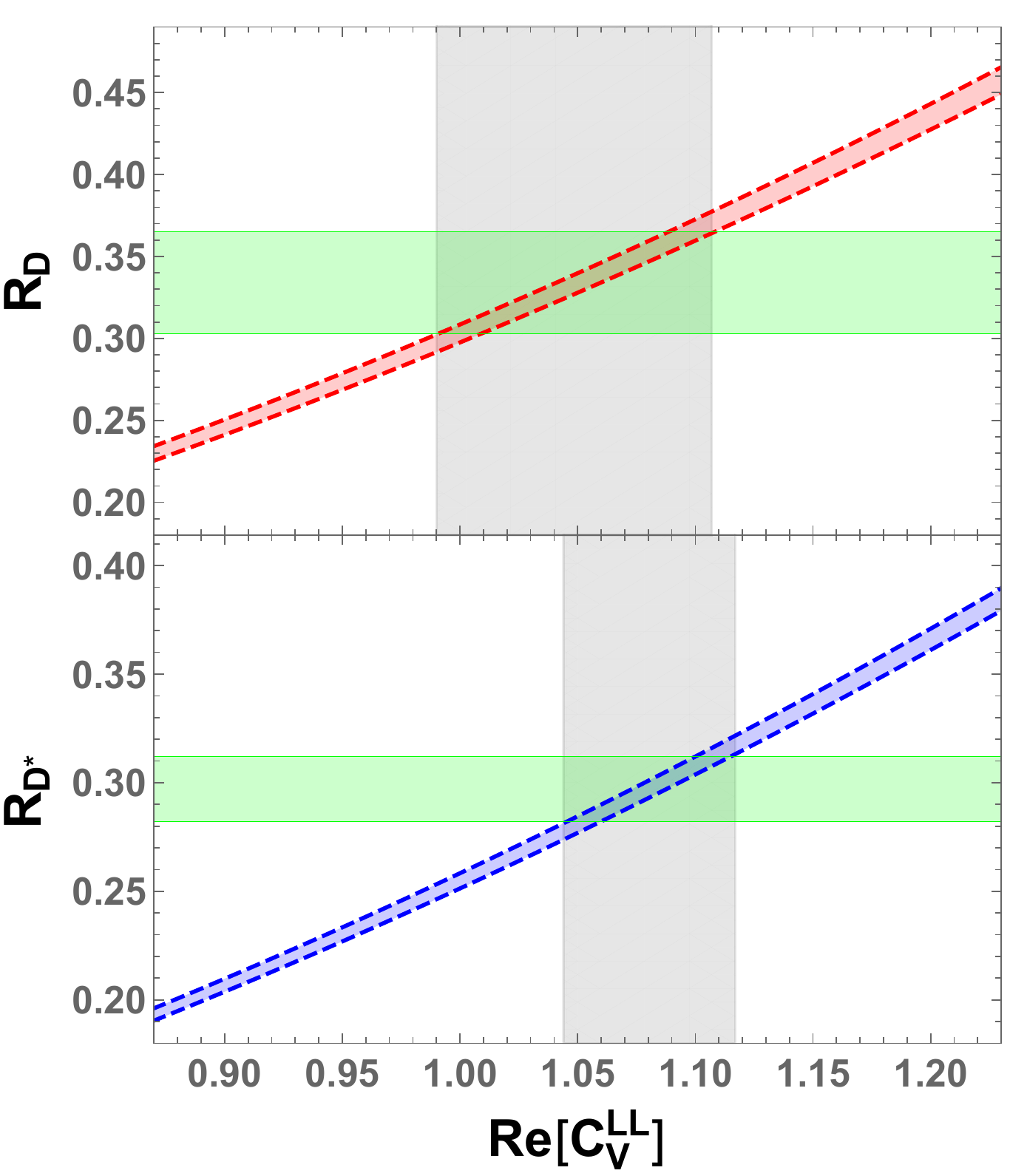} & \hspace*{-2.5mm} \includegraphics[scale=0.32]{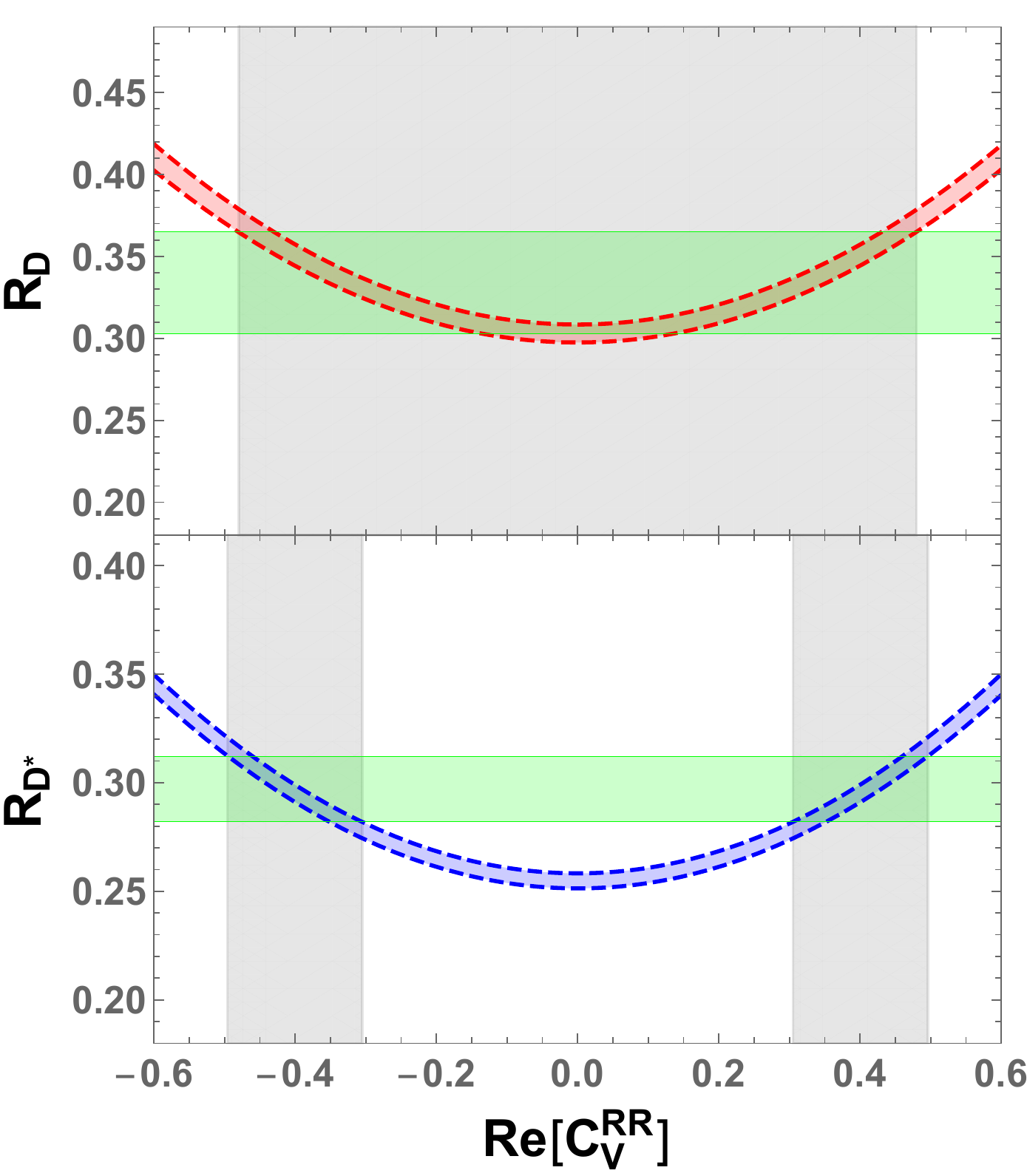}
\end{tabular}
\caption{\sf Variations of $R_D$ and $R_{D^*}$ against Re[$\cvll$]  and Re[$\cvrr$]. The green horizontal regions correspond to the
experimental $1\sigma$ average from table \ref{tab-exp} and the grey vertical regions correspond to the ranges of the WCs that produce $R_D$ and $R_{D^*}$
values within their $1\sigma$ experimental world average. Note that, $\cvll(\rm SM) = 1, \cvrr(\rm SM) = 0$.
\label{fig:cvll-rdrds}}
\end{figure}

In Fig.~\ref{fig:cvll-rdrds}, we show this for two WCs $\cvll$ and $\cvrr$ assuming them to be real.
%
%%%%%%%%%%%%%%%%%%%%%%%%%%%%%%
%%%%%%%%%%%%%%%%%%%%%%%%%%%%%%
%
It can be seen from the left panel that $\cvll=\cvll(\rm SM) =1$ is now at the edge of the $1\sigma$ allowed region for $R_D$. This is due to the fact
the the new experimental world-average for $R_D$ is now consistent with the SM expectation at $\sim 1\sigma$ level. So the anomaly is mostly
driven by $R_{D^*}$. In order to be consistent with both $R_D$ and $R_{D^*}$ simultaneously at the $1\sigma$ level, $\cvll$ has to be in the range
$\cvll:[1.045, 1.107]$. So there has not been a qualitative change in the situation after the new Belle measurement. Similarly, the allowed range for
$\cvrr$ now is $| \cvrr |:[0.305, 0.480]$. The lower edge of this range, $| \cvrr | = 0.305$, is now consistent with the $2\sigma$ upper bound $| \cvrr | = 0.32$
from the LHC $p \, p \to \tau \, \nu$
searches \cite{Greljo:2018tzh}\footnote{Note, however, that for $| \cvrr | = 0.305$, the value of $R_{D^{(*)}}$ is at the lower edge of the
experimental 1$\sigma$
allowed region. Moreover, the sensitivity of the current high-$p_T$ measurements is not enough to constrain the left-handed scenario $\cvll \approx 1.05$.
Thus, the right-handed scenario is statistically worse than the $\cvll$ solution.} (bound from LHC $p \, p \to \tau \, \nu + X$ searches was also studied in
\cite{Altmannshofer:2017poe, Iguro:2018fni}). Note that, both the WCs $\cvll$ and $\cvrr$ can be generated by a single
$U_1(3,1,2/3)$ Leptoquark mediator \cite{Alonso:2015sja,Barbieri:2015yvd,DiLuzio:2017vat,Azatov:2018kzb,Calibbi:2017qbu}.

\begin{figure}[!h!]
\centering
\begin{tabular}{cc}
\hspace*{-4mm}\includegraphics[scale=0.32]{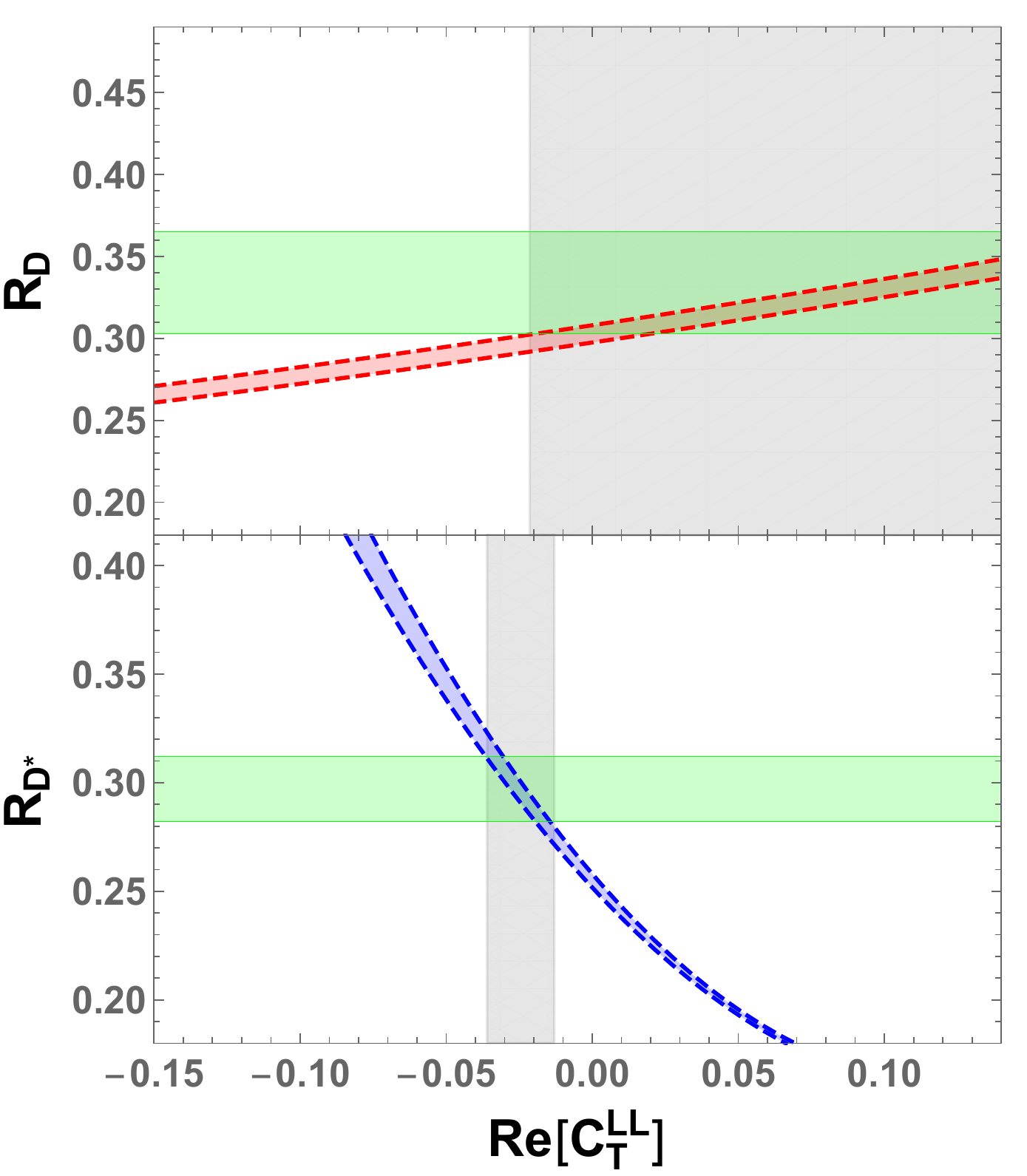}  & \hspace*{-2.5mm}  \includegraphics[scale=0.32]{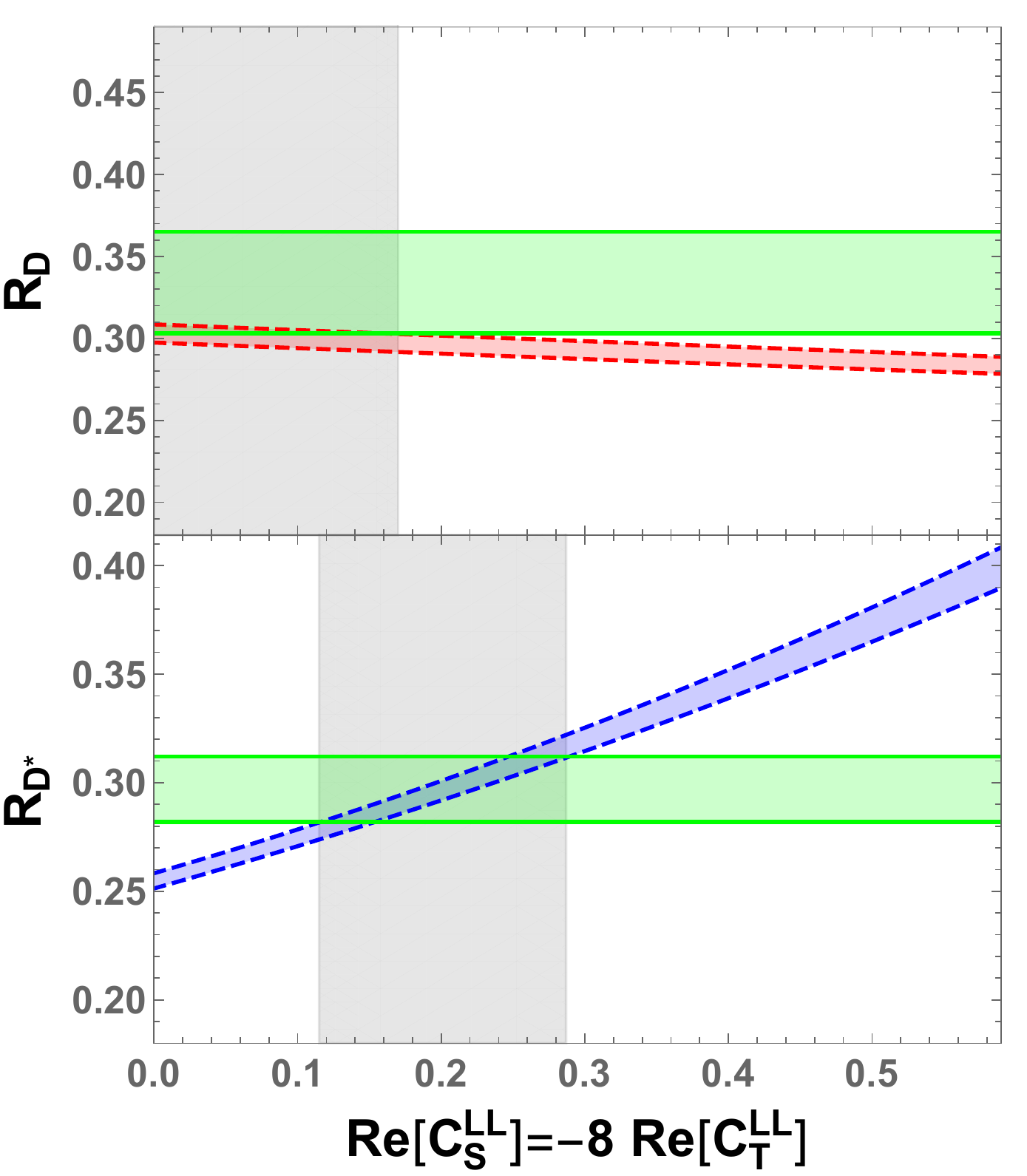}
\end{tabular}
\caption{\sf Variations of $R_D$ and $R_{D^*}$ against Re[$\ctll$] and $\rm Re[\csll]=-8 \rm Re[\ctll]$. \label{fig:ctll-csctll-rdrds}\\[0.4mm]}
\end{figure}

\begin{figure}[!h!]
\centering
\begin{tabular}{c}
\hspace*{-5mm} \includegraphics[scale=0.45]{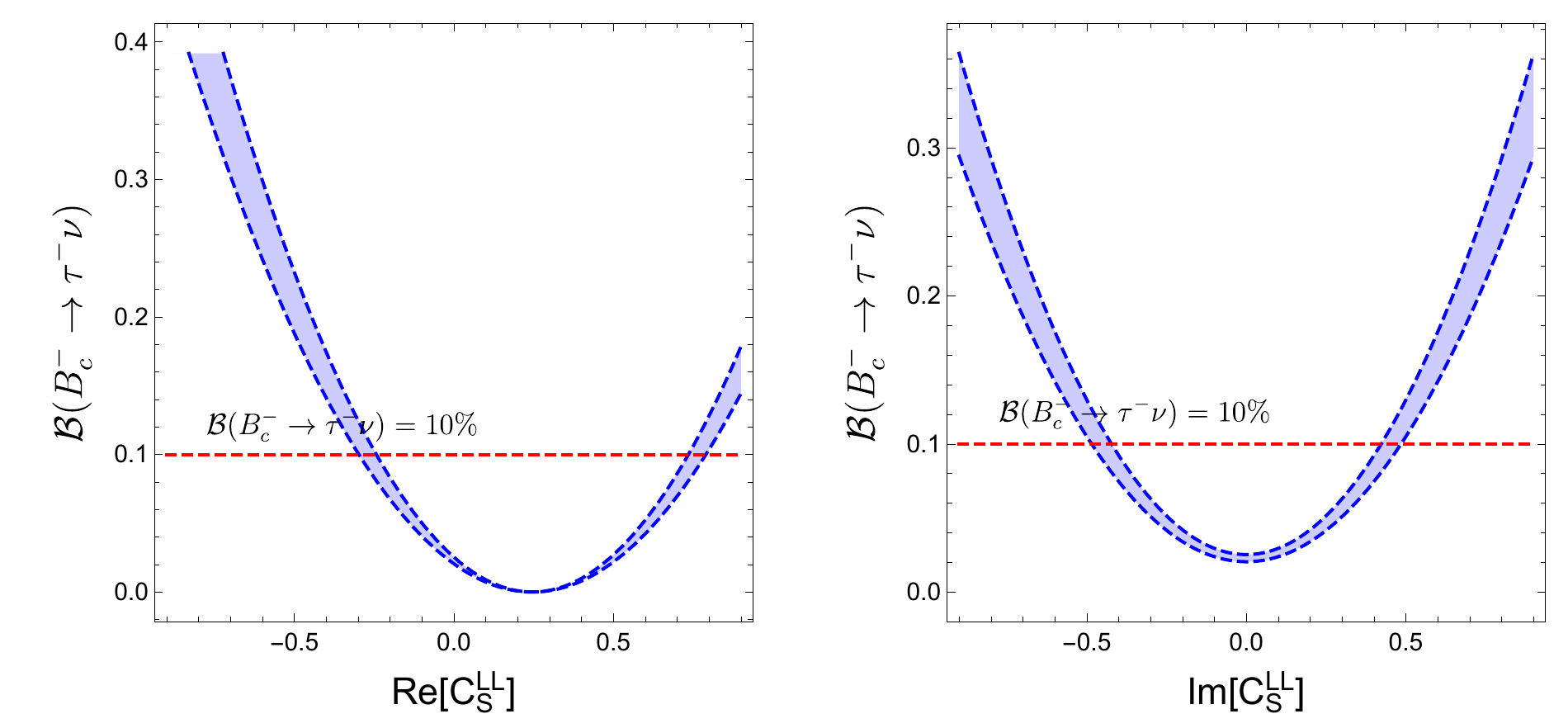} 
\end{tabular}
\caption{\sf Variation of ${\mathcal B}(B_c^- \to \tau^- \bar{\nu}_\tau)$ with respect to Re[$\csll$] and Im[$\csll$]. 
\label{fig:BcTauNu-branching}\\[0.4mm]}
\end{figure}

Variations of $R_D$ and $R_{D^*}$ with respect to $\ctll$ and $\csll=-8\ctll$ are shown in Fig.~\ref{fig:ctll-csctll-rdrds}.
It can be seen from the left panel of Fig.~\ref{fig:ctll-csctll-rdrds} that a simultaneous solution of $R_D$ and $R_{D^*}$ is possible for $\ctll$ in the
range $\ctll:[-0.021,-0.013]$. We remind the readers that the corresponding value of $\ctll$ before the recent Belle result was $\ctll \sim 0.35$
\cite{Bardhan:2016uhr,Azatov:2018knx} which was
strongly disfavoured both by the LHC $p \, p \to \tau \, \nu$ searches \cite{Aaboud:2018vgh,Sirunyan:2018lbg,Greljo:2018tzh} as well as the
measurement of $F_L^{D^*}$ \cite{mitp-talk}.
The new allowed range for $\ctll$, on the other hand, is completely safe. Thus, this has been a qualitative change after the new Belle measurement. 
%
%%%%%%%%%%%%%%%%%%%%%%%%%%%%%%
%%%%%%%%%%%%%%%%%%%%%%%%%%%%%%
%
The specific relation $\csll \approx -8\ctll$ (at the $m_b$ scale) shown on the right panel is interesting because it is generated by
a single $S_1(\bar{3},1,1/3)$ Leptoquark mediator \cite{Bauer:2015knc}.
The allowed range of the WC in this case is [0.113, 0.170] which, as can be seen from Fig.~\ref{fig:BcTauNu-branching},
produces ${\mathcal B}(B_c^- \to \tau^- \bar{\nu}_\tau)$ less then its SM value, and thus is completely safe.

%
%%%%%%%%%%%%%%%%%%%%%%%%%%%%%%
%%%%%%%%%%%%%%%%%%%%%%%%%%%%%%
%

\begin{figure}[!h!]
\centering
\begin{tabular}{cc}
\hspace*{-5mm} \includegraphics[scale=0.32]{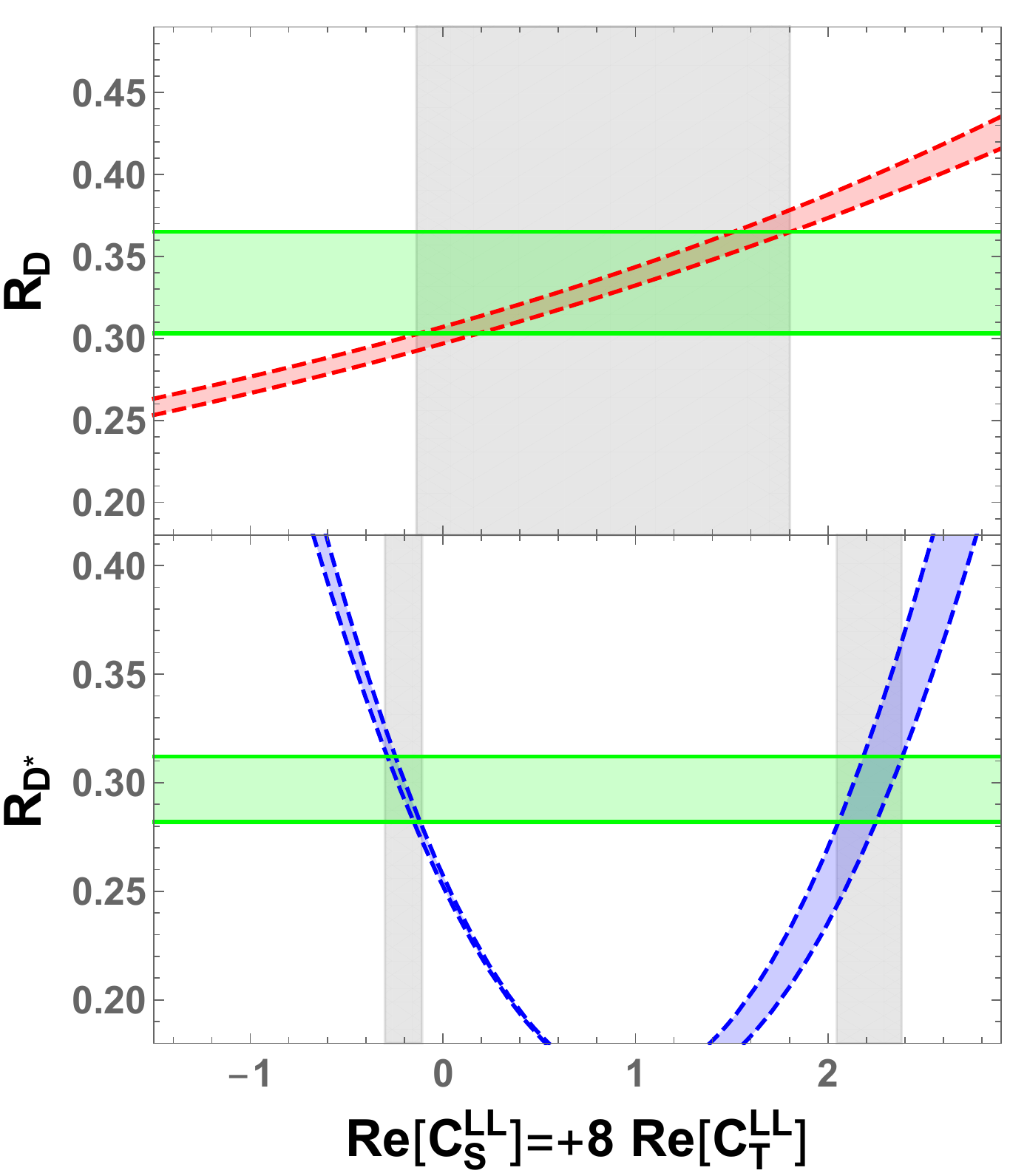} & \hspace*{-2.5mm} \includegraphics[scale=0.32]{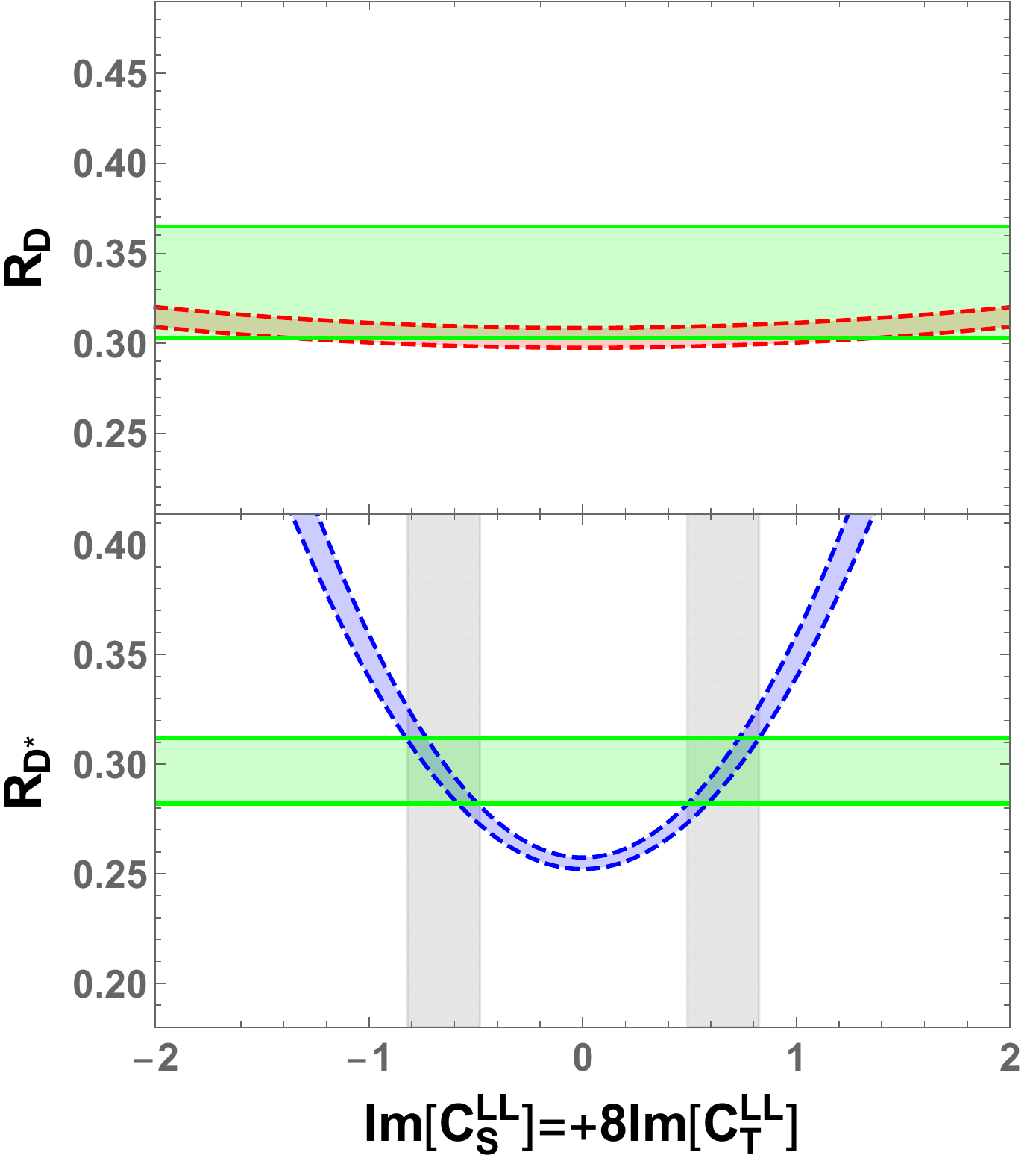}
\end{tabular}
\caption{\sf Variations of $R_D$ and $R_{D^*}$ against $\rm Re[\csll]= + 8 Re[\ctll]$ and $\rm Im[\csll]= + 8 Im[\ctll]$.
\label{fig:csctll-im-rdrds}\\[0.4mm]}
\end{figure}

Another single mediator solution that has been discussed in the literature is the so-called $R_2(3,2,7/6)$ Leptoquark \cite{Dorsner:2013tla,Becirevic:2018afm}.
which, contrary to the $S_1(\bar{3},1,1/3)$ Leptoquark mediator, generates $\csll \approx + 8\ctll$ (see the sign difference) at the $m_b$
scale\footnote{Note that, the relation $\csll = \pm 8 \ctll$ are approximately true only at the $m_b$ scale. It is obtained by QCD renormalization
group flow from the leptoquark matching scale ($\approx  {\rm few \, TeV}$) where the actual relations are $\csll = \pm 4 \ctll$.}.
In the left panel of Fig.~\ref{fig:csctll-im-rdrds}, we show this case assuming real values of the WCs. It can be seen that, the combination
$\rm Re[\csll] = + 8 Re[\ctll]$ at most can produce $R_D$ and $R_{D^*}$ at the lower edge of their $1\sigma$ experimental world-average if
a simultaneous solution is desired (for $\rm Re[\csll] = + 8 Re[\ctll] \approx -0.12$).
A much better description of the data is possible if imaginary WCs are assumed as shown in the right panel of Fig.~\ref{fig:csctll-im-rdrds}.
The case of imaginary WCs in this context was first discussed in \cite{Sakaki:2014sea}, and later also in \cite{Becirevic:2018afm,Blanke:2018yud, Iguro:2018vqb, Biswas:2018jun,Huang:2018nnq}.
%%%%%%%%%%%%%%%%%%%%%%%%%%%%%%
%%%%%%%%%%%%%%%%%%%%%%%%%%%%%%
%
In this case, one needs $\rm Im[\csll]= + 8 Im[\ctll]$ in the range [0.480, 0.820] which gives ${\mathcal B}(B_c^- \to \tau^- \bar{\nu}_\tau) > 10\%$,
see Fig.~\ref{fig:BcTauNu-branching}. However, the authors of Ref.~\cite{Akeroyd:2017mhr} claimed an upper bound of $10\%$ on this
branching ratio, arising from the LEP data taken on the $Z$ peak. 
Thus, the $\rm Im[\csll]= + 8 Im[\ctll]$ solution seems to be in slight tension if the $10\%$ upper bound is taken at  face value. 
While some authors \cite{Blanke:2018yud} expressed concerns about the validity of this bound, not much effort was made to estimate
as to how much this bound can be relaxed. We will discuss this in detail in the next section.

As the operator $\csrl$ alone cannot explain $R_D$ and $R_{D^*}$ simultaneously, we do not discuss it anymore.

Before concluding this section, we would like to make a couple of comments on the impact of $F_L^{D^*}$ and $P_\tau^{D^*}$
on the various scenarios. In all the scenarios explaining the $R_D$ and $R_D^*$ anomalies, the variation of $P_\tau^{D^*}$ 
is less than $\sim 2.5\%$ from the SM prediction. Unfortunately, this is also true about $F_L^{D^*}$, the only exception being
the $\rm Im[\csll]= 8 Im[\ctll]$ solution in which case the variation can be $5 -10\%$ below the SM. Thus, distinguishing the various
explanations by either $P_\tau^{D^*}$ or $F_L^{D^*}$ looks difficult at the moment.

%%
%%%%%%%%%%%%%%%%%%%%%%%%%%%%
%\begin{table}[h!]
%\begin{center}
%\begin{tabular}{|c|c|c|c|}
%\hline 
%                                                       &       $F_L^{D^*}$     &        $P_\tau^{D^*}$        &     $A_{FB}^{D^*}$  \\
%\hline
%$\rm SM$                                       &           -                     &               -                         &           -                      \\                                                     
%\hline
%$\cvll$                                            &           -                     &               -                         &           -                      \\ 
%\hline
%$\cvrr$                                           &           -                     &                -                        &           -                      \\ 
%\hline
%$\ctll$                                            &      0.430, 0.435      &    -0.502, -0.494               &    -0.046, -0.060      \\ 
%\hline
%$\rm Re[\csll] = - 8 \rm Re[\ctll]$   &      0.430, 0.434	    &   -0.501, -0.494              &  -0.058, -0.047        \\ 
%\hline
%$\rm Re[\csll] = + 8 \rm Re[\ctll]$  &     0.433, 0.447      &      -0.502, -0.496            &     -0.053, -0.104      \\ 
%\hline
%$\rm Im[\csll]  = + 8 \rm Im[\ctll]$  &      0.387, 0.420     &      -0.511, -0.510              &    -0.013, -0.054     \\ 
%\hline
%\end{tabular}
%\caption{\sf \label{tab-prediction}}
%\end{center} 
%\end{table}
%%%%%%%%%%%%%%%%%%%%%%%%%%%%%%%%%%%%%%%%%%%%%%%%%%%%%%%%%%%%%%%%%%%%% 
%%

%%%%%%%%%%%%%%%%%%%%%%%%%%%%%%%%%%%%%%%%%%%%%%%%%%%%%%%%%%%%
%\newpage
\vspace*{2mm}
{\bf LEP bound on ${\mathcal B}(B_c^- \to \tau^- \bar{\nu}_\tau)$: }
\vspace*{3mm}
%%%%%%%%%%%%%%%%%%%%%%%%%%%%%%%%%%%%%%%%%%%%%%%%%%%%%%%%%%%%

As mentioned in the previous section, the authors of \cite{Akeroyd:2017mhr} used the LEP data \cite{Acciarri:1996bv} collected at the $Z$ peak
to put an upper bound on the branching fraction of $B_c^- \to \tau^- \bar{\nu}_\tau$. As this constraint has potentially interesting
consequences for the $R_D$ and $R_{D^*}$ anomalies, in this section we will revisit it in detail.

In Ref.~\cite{Acciarri:1996bv}, the L3 collaboration obtained an upper bound
on the number of $B^- \to \tau^- \bar{\nu}_\tau$ events, $\mathcal{N}(B^- \to \tau^- \bar{\nu}_\tau) < 3.8$.
Based on this, they provided an upper bound
\begin{equation}
\mathcal{B}(B^- \to \tau \bar{\nu}_\tau) < 5.7 \times 10^{-4} \, \text{at} \, 90\% \, \text{C.L.} \label{eq:lep-1}
\end{equation}
As 
$
\mathcal{N}(B^- \to \tau^- \bar{\nu}_\tau) \propto f_{b \to B^-} \times \mathcal{B}(B^- \to \tau \bar{\nu}_\tau)
$
where, $f_{b \to B^-}$ is the inclusive probability that a $b$ quark hadronizes into a $B_c^-$ or  a $B_u^-$ meson,
and Ref.~\cite{Acciarri:1996bv} uses a value $f_{b \to B^-} = 0.382 \pm 0.025$, the bound in Eq.~\ref{eq:lep-1}
can be translated into the following bound 
\begin{equation}
f_{b \to B^-} \times \mathcal{B}(B^- \to \tau \bar{\nu}_\tau) < 2.035 \times 10^{-4}
\end{equation}

Separating the total number of events into those coming from $B_u^-$ and $B_c^-$ decays, we get 
\bal
& f_{b \to B_u^-} \, \mathcal{B}(B_u^- \to \tau^- \bar{\nu}_\tau)  +  f_{b \to B_c^-} \, \mathcal{B}(B_c^- \to \tau^- \bar{\nu}_\tau) \nn \\ 
& \hspace{5.2 cm}  < 2.035 \times 10^{-4}
\eal
This gives,
\bal
&\mathcal{B}(B_c^- \to \tau^- \bar{\nu}_\tau) <  \left(\frac{ 2.035 \times 10^{-4} }{f_{b \to B_u^-} \, \mathcal{B}(B_u^- \to \tau^- \bar{\nu}_\tau)} - 1 \right) \times \nn \\
& \hspace{4.2 cm} \frac{f_{b \to B_u^-}}{f_{b \to B_c^-}} \, \mathcal{B}(B_u^- \to \tau^- \bar{\nu}_\tau)
\label{eqn:BrC} 
\eal

The quantities $\mathcal{B}(B_u^- \to \tau^- \bar{\nu}_\tau)$ and $f_{b \to B_u^-}$ are known experimentally:
\bal
\mathcal{B}(B_u^- \to \tau^- \bar{\nu}_\tau) &= (1.06 \pm 0.20) \times 10^{-4} \, \text{\cite{Amhis:2016xyh,Tanabashi:2018oca}} \label{exp:BuTauNu}\\ 
%HFAG 2016 v3: pg 281
%PDG PRD2018: pg 705
f_{b \to B_u^-} &= 0.412 \pm  0.008  \, \text{\cite{Amhis:2016xyh,Tanabashi:2018oca}} \text{(LEP)} \label{LEP:fu}\\
f_{b \to B_u^-} &= 0.340 \pm 0.021  \,  \text{\cite{Amhis:2016xyh, Tanabashi:2018oca}} \text{(Tevatron)}  \label{TEV:fu}
%HFAG 2016 v3: pg 27
\eal
Note that, the hadronization fractions in $Z$ decays do not necessarily need to be identical to those in $p \, \bar{p}$ collisions
because of the different momentum distributions of the b-quark in these processes; in $p \, \bar{p}$ collisions, the
$b$ quarks have momenta close to $m_b$, rather than $\sim m_Z/2$ in $Z$ decays. In fact, CDF and LHCb collaborations
have reported evidence for a strong $p_T$ dependence of he $\Lambda_b^0$ fraction
\cite{Aaij:2011jp,Aaltonen:2008zd,Aaltonen:2008eu,Aaij:2014jyk}. The LHCb and the ATLAS collaborations
have also studied the $p_T$ dependence of $f_{b \to B_s}/f_{b \to B_d}$ \cite{Aaij:2013qqa,Aad:2015cda}, but the results are not conclusive yet.

Therefore, we use the measurement of $f_{b \to B_u^-}$ from LEP only and plot the upper bound on $\mathcal{B}(B_c^- \to \tau^- \bar{\nu}_\tau)$
as a function of $f_{b \to B_u^-}/f_{b \to B_c^-}$ in Fig.~\ref{LEP:fu}.
The upper bound $\mathcal{B}(B_c^- \to \tau^- \bar{\nu}_\tau) = 10\%$ corresponds to $f_{b \to B_u^-}/f_{b \to B_c^-} \approx 4 \times 10^{-3}$.

%
%%%%%%%%%%%%%%%%%%%%%%%%%%%%%%
\begin{figure}[!h!]
\centering
\begin{tabular}{c}
\includegraphics[scale=0.5]{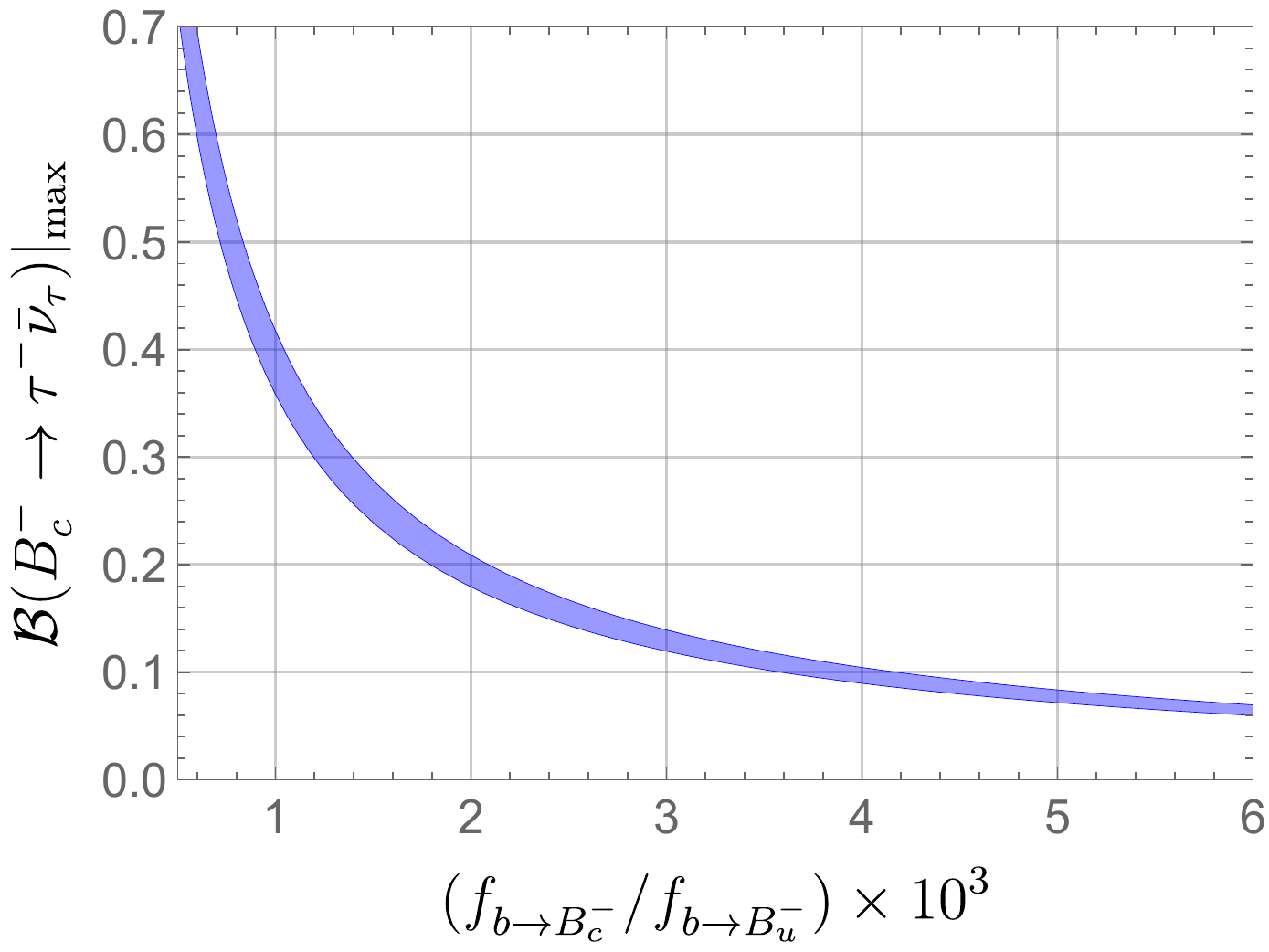}
\end{tabular}
\caption{\sf $\mathcal{B}(B_c^- \to \tau^- \bar{\nu}_\tau)|_{\rm max}$ as a function of $f_{b \to B_u^-}/f_{b \to B_c^-}$. The width of the plot 
corresponds to the uncertainties in Eq.~\eqref{exp:BuTauNu} and \eqref{LEP:fu}. \label{bctaunu-fcfu}\\[1mm]}
\end{figure}
%%%%%%%%%%%%%%%%%%%%%%%%%%%%%%
%

In order to find a real upper bound on $\mathcal{B}(B_c^- \to \tau^- \bar{\nu}_\tau)$ we need to know the value
of $f_{b \to B_u^-}/f_{b \to B_c^-}$, or at least a lower bound on $f_{b \to B_u^-}/f_{b \to B_c^-}$. Moreover, we need to
know $f_{b \to B_u^-}/f_{b \to B_c^-}$ at LEP, and with the exact kinematical cuts used in \cite{Acciarri:1996bv}.

Ref.~\cite{Akeroyd:2017mhr} tries to find the ratio $f_{b \to B_u^-}/f_{b \to B_c^-}$ from measurements of
$R_{\pi^+/K^+}$ and $R_{\pi^+/\mu^+}$ defined as
\bal
R_{\pi^+/K^+} &= \frac{f_{\bar{b} \to B_c^+}}{f_{\bar{b} \to B_u^+}}\,
\frac{\mathcal{B}\left(B_c^+ \to J/\psi \, \pi^+ \right)}{\mathcal{B}\left(B_u^+ \to J/\psi \, K^+ \right)} \\
R_{\pi^+/\mu^+} &= \frac{\mathcal{B}\left(B_c^+ \to J/\psi \, \pi^+ \right)}{\mathcal{B}\left(B_c^+ \to J/\psi \, \mu^+ \, \nu\right)} \, .
\eal

It then follows that
\begin{eqnarray}
\frac{f_{\bar{b} \to B_c^+}}{f_{\bar{b} \to B_u^+}}\, \frac{\mathcal{B}\left(B_c^+ \to J/\psi \, \mu^+ \, \nu_\mu \right)}{\mathcal{B}\left(B_u^+ \to J/\psi \, K^+ \right)}
= \frac{R_{\pi^+/K^+}}{R_{\pi^+/\mu^+}} \\
\Rightarrow \, \frac{f_{\bar{b} \to B_c^+}}{f_{\bar{b} \to B_u^+}} 
= \frac{\mathcal{B}\left(B_u^+ \to J/\psi \, K^+\right)}{\mathcal{B}\left(B_c^+ \to J/\psi \, \mu^+ \, \nu_\mu\right)}     \frac{R_{\pi^+/K^+}}{R_{\pi^+/\mu^+}}  \, .
\end{eqnarray}

Using 
\bal
R_{\pi^+/\mu^+} & = 0.0469 \pm 0.0054 \, \text{\cite{Aaij:2014jxa}} \\
R_{\pi^+/K^+}& =^{^{\hspace{-4mm} \rm LHCb}}  (0.683 \pm 0.02) \times 10^{-2} \text{\cite{Aaij:2014ija}}  \label{fcfu-lhcb-0}\\ 
                       & =^{^{\hspace{-4mm} \rm CMS}}  (0.48 \pm 0.08) \times 10^{-2} \text{\cite{Khachatryan:2014nfa}}  \label{fcfu-cms-0} \\ 
\mathcal{B}\left(B_u^- \to J/\psi \, K^-\right) & =  (9.99 \pm 0.36) \times 10^{-4}  \text{\cite{Amhis:2016xyh}}
%https://hflav-eos.web.cern.ch/hflav-eos/b2charm/Spring_2018/index.html
\eal
we get,
\bal
\frac{f_{\bar{b} \to B_c^+}}{f_{\bar{b} \to B_u^+}} 
&= \frac{(1.22 - 1.75) \times 10^{-4}}{\mathcal{B}\left(B_c^+ \to J/\psi \, \mu^+ \, \nu_\mu\right)}  (\text{using \cite{Aaij:2014ija}}) \label{fcfu-lhcb} \\
\frac{f_{\bar{b} \to B_c^+}}{f_{\bar{b} \to B_u^+}}
& = \frac{(0.74 - 1.40) \times 10^{-4}}{\mathcal{B}\left(B_c^+ \to J/\psi \, \mu^+ \, \nu_\mu\right)}  (\text{using \cite{Khachatryan:2014nfa}})  \label{fcfu-cms}
\eal
As the LHCb and CMS measurements of $R_{\pi^+/K^+}$ are about $2.5\sigma$ away from each other, we consider them separately
and do not use their average.
Moreover, while the LHCb Collaboration uses the cuts $0 < p_T (B_c^+), \, p_T (B_u^+) < 20 \ {\rm GeV}$ and $2.0  <  \eta < 4.5$ in their
analysis (at $\sqrt{s} = 8 \ {\rm TeV}$), the CMS Collaboration uses $p_T (B_c^+), \, p_T (B_u^+) > 15 \ {\rm GeV}$ and $|\eta| < 1.6$
(at $\sqrt{s} = 7 \ {\rm TeV}$). Thus the discrepancy could be due to the dependence of $f_{\bar{b} \to B_c^+}/f_{\bar{b} \to B_u^+}$
on kinematics.

\begin{figure}[!h!]
\centering
\begin{tabular}{cc}
\hspace*{-4mm} \includegraphics[scale=0.35]{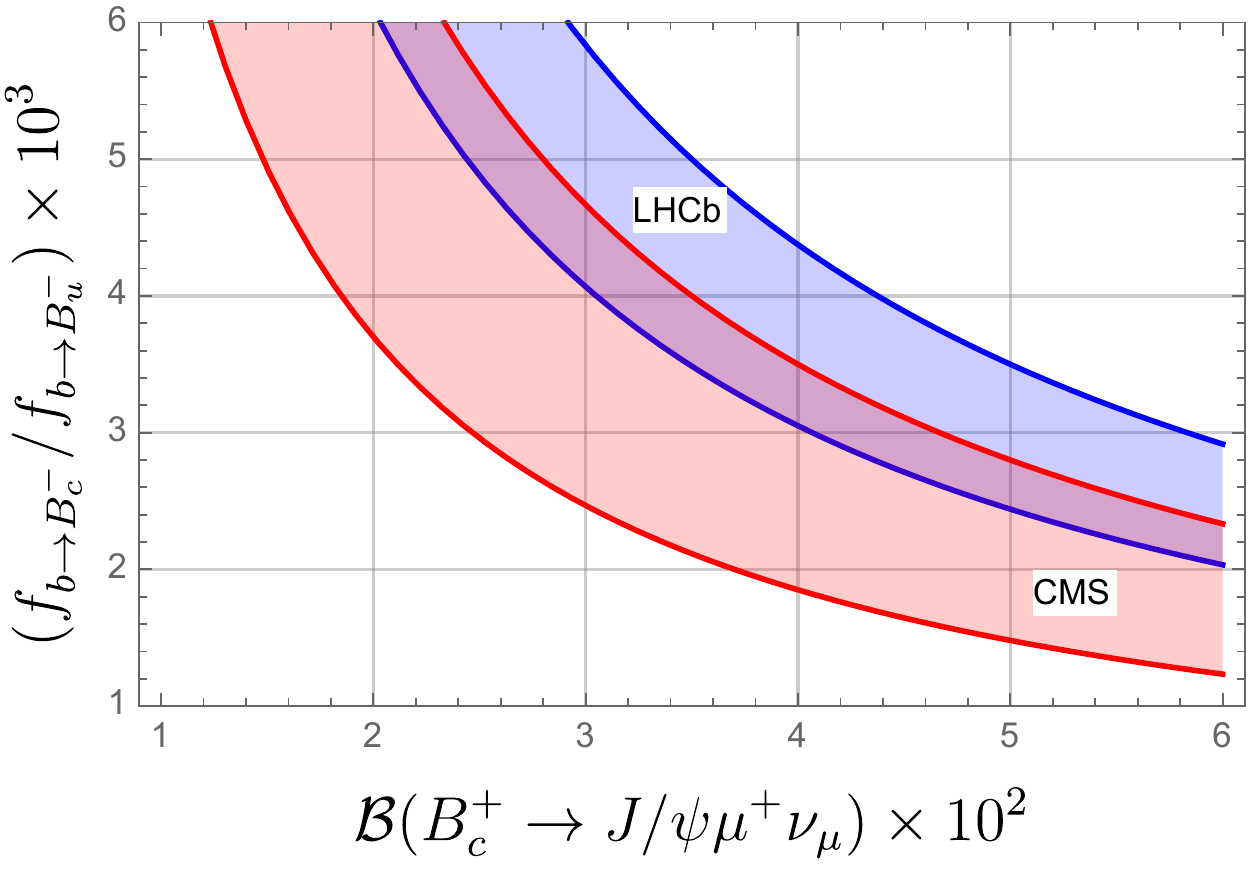} & \hspace*{-2mm} \includegraphics[scale=0.35]{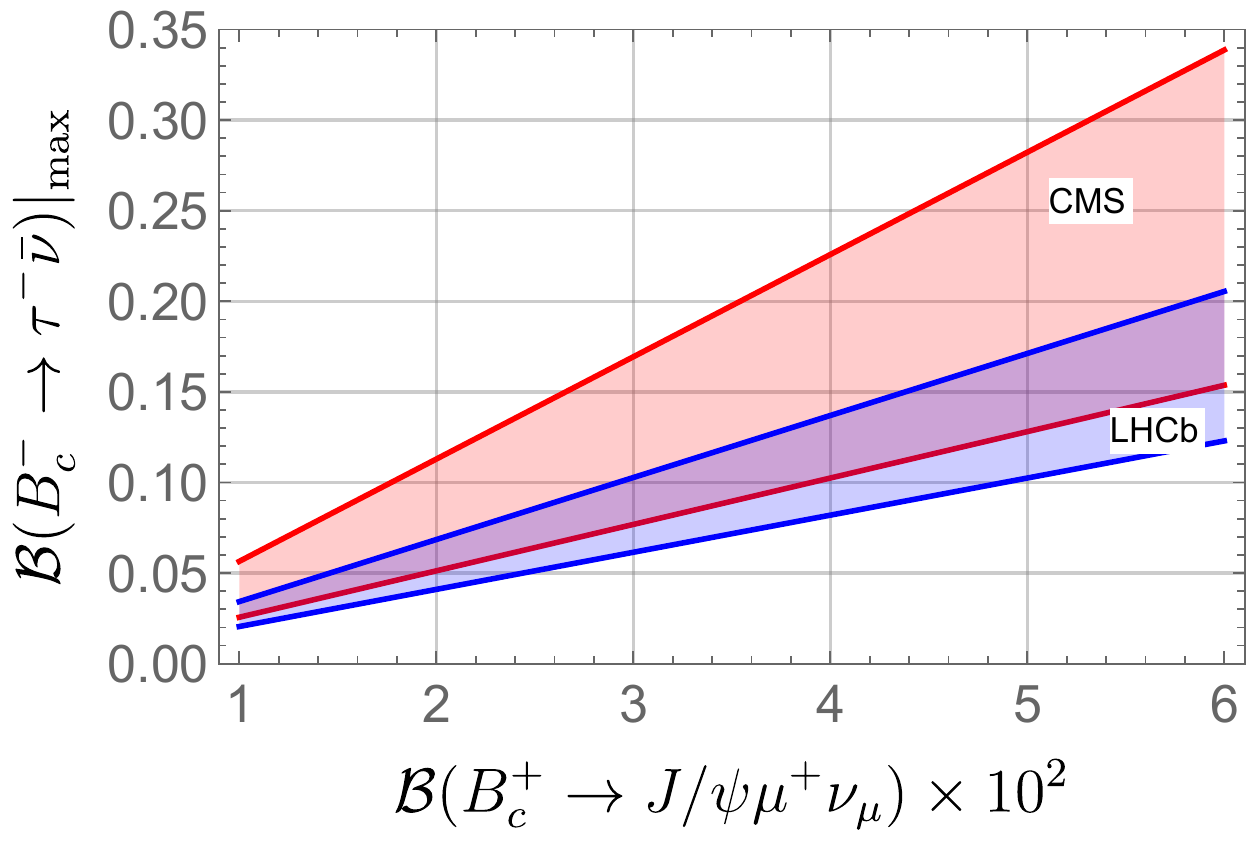}
\end{tabular}
\caption{\sf Variations of $f_{\bar{b} \to B_c^+}/f_{\bar{b} \to B_u^+}$ and $\mathcal{B}(B_c^- \to \tau^- \bar{\nu}_\tau)|_{\rm max}$ with respect to
${\mathcal{B}\left(B_c^+ \to J/\psi \, \mu^+ \, \nu_\mu\right)}$.
\label{fcfu-Jpsi}}
\end{figure}

Plugging Eqs.~\ref{fcfu-lhcb} and \ref{fcfu-cms} into Eq.~\ref{eqn:BrC}, one can obtain a bound on $\mathcal{B}(B_c^- \to \tau^- \bar{\nu}_\tau)$
directly as a function of ${\mathcal{B}\left(B_c^+ \to J/\psi \, \mu^+ \, \nu_\mu\right)}$. This is shown in the right panel of Fig.~\ref{fcfu-Jpsi}.
%
%%%%%%%%%%%%%%%%%%%%%%%%%%%%%%
%%%%%%%%%%%%%%%%%%%%%%%%%%%%%%
%

Using ${\mathcal{B}\left(B_c^+ \to J/\psi \, \mu^+ \, \nu_\mu\right)} \leq 2.5 \times 10^{-2}$, as used in \cite{Akeroyd:2017mhr}, we get
$f_{\bar{b} \to B_c^+}/f_{\bar{b} \to B_u^+} \gtrsim 3 \times 10^{-3}$ and $\mathcal{B}(B_c^- \to \tau^- \bar{\nu}_\tau) \lesssim 14\%$
from the CMS data, the latter being similar but slightly weaker than \cite{Akeroyd:2017mhr}.

We would like to make two comments at this stage:
\begin{itemize}
\item The bound on $\mathcal{B}(B_c^- \to \tau^- \bar{\nu}_\tau)$ depends linearly on ${\mathcal{B}\left(B_c^+ \to J/\psi \, \mu^+ \, \nu_\mu\right)}$. 
As ${\mathcal{B}\left(B_c^+ \to J/\psi \, \mu^+ \, \nu_\mu\right)}$ has not yet been measured, a model independent bound is not possible. 
Moreover, even the SM calculation, and in particular the uncertainty, is not fully under control at the moment. Thus, a precise bound on
$\mathcal{B}(B_c^- \to \tau^- \bar{\nu}_\tau)$ cannot be obtained currently. 

\item Even in the presence of better information on ${\mathcal{B}\left(B_c^+ \to J/\psi \, \mu^+ \, \nu_\mu\right)}$, Eqs.~\eqref{fcfu-lhcb} and \eqref{fcfu-cms}
provide values of  $f_{\bar{b} \to B_c^+}/f_{\bar{b} \to B_u^+}$ at the LHC and for the specific kinematic regions used in \cite{Aaij:2014ija}
and \cite{Khachatryan:2014nfa}. As discussed before, the value of $f_{\bar{b} \to B_c^+}/f_{\bar{b} \to B_u^+}$ at LEP may be different from the above
because of 1) larger average $p_T$ of the b-mesons produced at LEP 2) $b \bar{b}$ pairs produced at LEP are in the colour singlet state contrary to
most of the $b \bar{b}$ pairs produced at the LHC which are in the colour octet state.
\end{itemize}

In view of the above, we try to estimate the ratio $f_{\bar{b} \to B_c^+}/f_{\bar{b} \to B_u^+}$ at LEP using the event generator
Pythia8 \cite{Sjostrand:2006za,Sjostrand:2014zea} which has Hadronization model tuned to provide a good description of the available
experimental data.
The results are shown in Table.~\ref{tab-pythia}. In each of the cases presented in Table.~\ref{tab-pythia}, we have generated 1 million events
in order to reduce the statistical uncertainty. In Case-I, we have used the same $p_T$ and $\eta$ cuts as in \cite{Khachatryan:2014nfa}, and we get
a value $f_{b \to B_c^-}/f_{b \to B_u^-} = 1.06 \times 10^{-3}$ which is much smaller than  $f_{b \to B_c^-}/f_{b \to B_u^-} = 3 \times 10^{-3}$ which
was used to obtain a bound $\mathcal{B}(B_c^- \to \tau^- \bar{\nu}_\tau) \leq 10\%$. Note that, from Eq.~\ref{fcfu-cms},
$f_{b \to B_c^-}/f_{b \to B_u^-} = 1.06 \times 10^{-3}$ would correspond to ${\mathcal{B}\left(B_c^+ \to J/\psi \, \mu^+ \, \nu_\mu\right)} \approx 6 \times 10^{-2}$
(see the left panel of Fig.~\ref{fcfu-Jpsi}) which is much larger than the values considered in \cite{Akeroyd:2017mhr}. 
%
%%%%%%%%%%%%%%%%%%%%%%%%%%%
\begin{table}[h!]
\begin{center}
\begin{tabular}{|c|c|c|c|c|}
\hline 
                  &      &     $f_{b \to B_u^-}$  &  $f_{b \to B_c^-}$  &  $\dfrac{f_{b \to B_c^-}}{f_{b \to B_u^-}}$  \\
\hline
          &     LHC 7 TeV                                                       &              &               &              \\ 
  I        &    $p_T (B_c^+, B_u^+) > 15 \ {\rm GeV}$    &    0.255       & $ 2.7 \times 10^{-4}$     &   $1.06 \times 10^{-3}$     \\     
          &    $ |\eta| < 1.6$                                                  &                &               &                \\     
\hline
          &    LHC 7 TeV                                                       &              &               &              \\ 
 II         &    $p_T (B_c^+, B_u^+) < 15 \ {\rm GeV}$    &    0.301     &  $ 5.7 \times 10^{-4}$   &     $1.89 \times 10^{-3}$      \\     
          &    $ |\eta| < 1.6$                                                  &                &               &                \\     
\hline
          &    LHC 7 TeV                                                       &              &               &              \\ 
          &    $q \bar{q}  \to Z \to b \bar{b}$   only                &              &               &              \\ 
 III       &    $p_T (B_c^+, B_u^+) > 15 \ {\rm GeV}$        &      0.374       &  $ 4.1 \times 10^{-4}$   &    $1.09 \times 10^{-3}$             \\     
          &    $ |\eta| < 1.6$                                                  &                &               &                \\     
\hline
          &    LHC 7 TeV                                                       &              &               &              \\ 
          &    $g g \to b \bar{b}$ and $q \bar{q} \to g \to b \bar{b}$                        &              &               &              \\ 
 IV        &    $p_T (B_c^+, B_u^+) > 15 \ {\rm GeV}$    &   0.255      &  $ 2.5 \times 10^{-4}$      &    $0.98 \times 10^{-3}$              \\     
          &    $ |\eta| < 1.6$                                               &                &               &                \\     
\hline
V          &    LEP (at the Z peak)                                     &    0.42            &   $ 4.5 \times 10^{-4}$    &       $1.07 \times 10^{-3}$         \\ 
\hline
\end{tabular}
\caption{\sf Hadronization fractions calculated from Pythia8. \label{tab-pythia}}
\end{center} 
\end{table}
%%%%%%%%%%%%%%%%%%%%%%%%%%%%%%%%%%%%%%%%%%%%%%%%%%%%%%%%%%%%%%%%%%%% 
%
In the third row of Table.~\ref{tab-pythia}, we changed the $p_T$ cut to $p_T < 15 \text{GeV}$ in order to check the $p_T$ dependence
of the Hadronization fractions. In this case, we get ${f_{b \to B_c^-}}/{f_{b \to B_u^-}} = 1.89 \times 10^{-3}$ which is considerably larger than
that in Case-I. This is consistent with the general findings in \cite{Aaij:2011jp,Aaltonen:2008zd,Aaltonen:2008eu,Aaij:2014jyk, Aaij:2013qqa,Aad:2015cda}
and confirms that the measurement of  ${f_{b \to B_c^-}}/{f_{b \to B_u^-}}$ from LHCb (Eq.~\ref{fcfu-lhcb-0} and \ref{fcfu-lhcb}) which uses
$p_T (B_c^+), \, p_T (B_u^+) < 20 \ {\rm GeV}$ is indeed not expected to be the same as that measured in CMS (Eq.~\ref{fcfu-cms-0} and \ref{fcfu-cms})
which used $p_T (B_c^+), \, p_T (B_u^+) > 15 \ {\rm GeV}$.
In rows 4 and 5 of Table.~\ref{tab-pythia}, we considered $b \, \bar{b}$ production through only Z boson (produced $b \bar{b}$ are in QCD singlet state) and
through only QCD interactions (produced $b \bar{b}$ are in QCD triplet state) respectively. We observed only $\sim 10\%$ variation in the
${f_{b \to B_c^-}}/{f_{b \to B_u^-}}$
between these two cases.

Finally, at the Z peak, we obtain $f_{b \to B_u^-} = 0.42$, $f_{\bar{b} \to B_s} = 0.094$ (not shown in the table), and
${f_{b \to B_c^-}}/{f_{b \to B_u^-}} =1.07 \times 10^{-3}$, the first two numbers being consistent with their experimental
measurements \cite{Amhis:2016xyh,Tanabashi:2018oca}. Using the number ${f_{b \to B_c^-}}/{f_{b \to B_u^-}} =1.07 \times 10^{-3}$, from
Fig.~\ref{bctaunu-fcfu}, we get
\bal
\mathcal{B}(B_c^- \to \tau^- \bar{\nu}_\tau) \leq 39\% \, .
\eal
We warn the readers that this bound should only be taken as an estimate because, after all, Pythia only uses a Hadronization model
adjusted to describe a large amount of available experimental data well (as we saw, indeed it reproduced the correct values for $f_{b \to B_u^-}$ and
$f_{\bar{b} \to B_s}$), and the value of $f_{b \to B_c^-}$ obtained from Pythia is neither based on any first principle calculation nor on direct
experimental data.

%%%%%%%%%%%%%%%%%%%%%%%%%%%%%%%%%%%%%%%%%%%%%%%%%%%%%%%%%%%%
%\newpage
%\vspace*{2mm}
%{\bf Summary: }
\vspace*{3mm}
%%%%%%%%%%%%%%%%%%%%%%%%%%%%%%%%%%%%%%%%%%%%%%%%%%%%%%%%%%%%

To summarise, in this short note, we have shown that 
\begin{itemize} 
\item  the recent Belle results on $R_D$ and $R_{D^*}$  have interesting implications on the various possible EFT 
explanations of the data. The most important being that the pure tensor explanation is now completely allowed both by
the measurement of $F_L^{D^*}$ and the high-$p_T$ $p \, p \to \tau \, \nu$ searches by ATLAS and CMS.

\item the solution in terms of a pure right-chiral vector current (involving right-chiral neutrinos) has now moved  into the
$2\sigma$ allowed range of the LHC $p \, p \to \tau \, \nu$ searches.

\item the upper bound on the branching fraction of $B_c^- \to \tau^- \bar{\nu}_\tau$ from the LEP data is much weaker than
the bound $10\%$ used in the recent literature. Our estimate of this bound, based on the Hadronization model implemented
in Pythia8, is approximately $40\%$. This bound, while being independently important, may also have interesting implications
on the various scalar-pseudoscalar explanations of the $R_D$ and $R_{D^*}$ data. 
\end{itemize}

{\bf Acknowledgement } \\
The research of DB was supported in part by the Israel Science Foundation (grant no. 780/17) and by the Kreitman Foundation
Post-Doctoral Fellowship. 
DG would like to acknowledge support through Ramanujan Fellowships of the Department of Science and Technology, Government of India.

\vspace*{3mm}

%%%%%%%%%%%%%%%%%%%%%%%%%%%%%%%%%%%%%%%%%%%%%%%%%%%%%%%%%%%%%%%%%%%%%%%%%%%%%%%

%%%%%%%%%%%%%%%%%%%%%%%%%%%%%%%%%%%%%%%%%%%%%%%%%%%%%%%%%%%%%%%%%%%%%%%%%%%%%%%
%\clearpage
%\newpage
%\newpage
%\vspace*{2mm}
{\bf References }
\vspace*{2mm}
%\bibliographystyle{bibtex/JHEP}
%\bibliography{bibtex/references}
\providecommand{\href}[2]{#2}\begingroup\raggedright\endgroup
%%%%%%%%%%%%%%%%%%
\end{document}